%% file: ms.tex
\documentclass[12pt,preprint]{aastex}

\received{}
\revised{}
\accepted{}

\shorttitle{Disk Galaxies in the SDSS}
\shortauthors{Unterborn \& Ryden}

\begin{document}
\title{Inclination-Dependent Extinction Effects in Disk
Galaxies in the Sloan Digital Sky Survey}
\author{Cayman T. Unterborn \& Barbara S. Ryden\altaffilmark{1}}
\affil{Department of Astronomy, The Ohio State University,
Columbus, OH 43210}
\altaffiltext{1}{Center for Cosmology \& Astro-Particle Physics,
The Ohio State University, Columbus, OH 43210}
\email{unterborn.1@osu.edu, ryden@astronomy.ohio-state.edu}

\begin{abstract}
We analyze the absolute magnitude ($M_r$) and color
($u-r$) of low redshift ($z < 0.06$) galaxies in the
Sloan Digital Sky Survey Data Release 6. Galaxies
with nearly exponential profiles (Sloan parameter
${\rm fracDeV} < 0.1$) all fall on the blue sequence
of the color -- magnitude diagram; if, in addition,
these exponential galaxies have $M_r < -19$, they
show a dependence of $u-r$ color on apparent
axis ratio $q$ expected for a dusty disk galaxy.
By fitting luminosity functions for exponential
galaxies with different values of $q$, we find
that the dimming is well described by the relation
$\Delta M_r = 1.27 ( \log q )^2$, rather than the
$\Delta M \propto \log q$ law that is frequently
assumed. When the absolute magnitudes of bright
exponential galaxies are corrected to their
``face-on'' value, $M_r^f = M_r - \Delta M_r$,
the average $u-r$ color is linearly dependent
on $M_r^f$ for a given value of $q$. Nearly face-on
exponential galaxies ($q > 0.9$) have a shallow dependence
of mean $u-r$ color on $M_r^f$ (0.096 magnitudes redder
for every magnitude brighter); by comparison, nearly edge-on
exponential galaxies ($q < 0.3$) are 0.265 magnitudes redder
for every magnitude brighter.
When the dimming law $\Delta M_r \propto ( \log q )^2$ 
is used to create an inclination-corrected sample of
bright exponential galaxies, their apparent shapes
are confirmed to be consistent with a distribution
of mildly non-circular disks, with median short-to-long
axis ratio $\gamma \approx 0.22$ and median
disk ellipticity $\epsilon \approx 0.08$.

\end{abstract}


\section{INTRODUCTION}
\label{sec-intro}

Luminous galaxies ($M_r < -18$ or so) can be coarsely divided
into two classes, conventionally labeled ``early-type'' and
``late-type''. Early-type galaxies have redder stellar
populations and a scarcity of interstellar gas and dust.
The majority of luminous
early-type galaxies are elliptical galaxies,
characterized by smooth isophotes and concentrated
light profiles, well described by a
\citet{dV48} profile: $\log I \propto - r^{1/4}$.
Highly luminous elliptical galaxies tend to be mildly triaxial
ellipsoids (as opposed to perfectly oblate spheroids);
their intrinsic short-to-long axis ratio is
typically $c / a \sim 0.7$ \citep{ry92,vi05}.

Late-type galaxies have blue stellar populations
and relatively large amounts of interstellar gas and dust.
The majority of luminous late-type galaxies are spiral
galaxies, characterized by spiral structure within
flattened disks. The disk light profile is generally
well described by an exponential profile: $\log I \propto -r$.
Luminous spiral galaxies are mildly elliptical (as opposed
to perfectly circular) when seen face-on; their intrinsic
short-to-long axis ratio is color-dependent, but at
visible wavelengths is typically $c / a \sim 0.25$
\citep{bi81,gr85,la92,fa93,ry04,ry06}.

In a color-magnitude (CM) diagram, if the color index is
chosen correctly, the early-type and late-type galaxies
manifest themselves as a ``red sequence'' and a ``blue
sequence'', respectively. In the Sloan Digital Sky Survey (SDSS),
the distribution of $u-r$ colors for low-redshift
galaxies is bimodal; \citet{st01} find an optimal
color separator of $u-r = 2.22$, when color alone
is used as a discriminator between early-type and
late-type galaxies. Using a full CM diagram,
the color separator is found to be dependent on
$M_r$ \citep{ba04}, with the color separator ranging
from $u-r \approx 2.3$ for galaxies with $M_r < -21$
to $u-r \approx 1.8$ for galaxies with $M_r > -18$.

However, a clean separation between early-type and
late-type galaxies using color and absolute magnitude
information alone is impossible; the red sequence
and blue sequence overlap in a CM diagram. This overlap
results partly from the fact that the color and apparent
magnitude of spiral galaxies are inclination dependent.
Since spiral galaxies are disk-shaped and contain dust,
an edge-on spiral is both redder and fainter than the
same spiral would be if seen face-on. Thus, as noted
by \citet{al02}, a sample chosen solely by the
color criterion $u-r \geq 2.22$ will contain dust-reddened
edge-on spirals as well as intrinsically red ellipticals.

The overlap between the red sequence and blue sequence would
be reduced if we could perform an inclination correction
on the colors and apparent magnitudes of spiral galaxies;
that is, if we could convert observed apparent magnitudes
into what they would be if the spiral galaxy were face-on.
Since pivoting galaxies so that we can see them face-on
is an impracticable task, we will take a statistical
approach to finding the average dimming ($\Delta M_r$)
and reddening ($\Delta (u-r)$) of a spiral
galaxy as a function of its inclination $i$.

In addition to allowing a cleaner separation between
early-type and late-type galaxies in the CM diagram,
a statistical correction for dimming and reddening
has other practical uses. For instance, a flux-limited
survey will undersample edge-on spirals with respect
to face-on spirals; a galaxy that is just above the
flux limit when it is face on would fall below the
limit if it were edge on. With a knowledge of $\Delta M_r$ 
as a function of inclination, it is possible to
create an inclination-corrected flux-limited sample,
by retaining only those galaxies that would still
be above the flux limit if they were tilted to be
seen edge on. Since the standard technique for
finding the distribution of intrinsic axis ratios
of galaxies assumes a random distribution of inclinations
(see, for instance, \citep{vi05} and references therein),
such an inclination-corrected sample is essential for
determining the true distribution of disk flattening.
Without the inclination correction, the scarcity of
edge-on galaxies would lead to an overestimate of the
typical disk thickness.

In section~\ref{sec-data}, we describe how
we select a sample of SDSS galaxies with exponential
profiles; this gives us a population of disk-dominated
spiral galaxies. The apparent axis ratio $q$ of the
25 mag arcsec$^{-2}$ isophote is chosen as our surrogate
for the inclination of a galaxy. In section~\ref{sec-analysis},
we examine the luminosity function of the SDSS exponential
galaxies as a function of $q$. By shifting the luminosity
functions until their high-luminosity cutoffs align, we can 
estimate the dimming $\Delta M_r$ of the cutoff as a function of $q$.
By brightening each galaxy by the $\Delta M_r$ appropriate
to its observed value of $q$, we can find its approximate
``face-on'' absolute magnitude $M_r^f$. We then provide
linear fits for the mean $u-r$ color as a function of
$M_r^f$ for different ranges of $q$. In section~\ref{sec-shapes},
we create an inclination-corrected flux-limited sample,
which we then use to find the distribution of intrinsic
short-to-long axis ratios of the SDSS exponential galaxies,
confirming that they are, in fact, disks, as our analysis
assumed all along. Finally, in section~\ref{sec-disc}, we
provide a brief discussion of what the form of $\Delta M_r$
as a function of $q$ and of $\Delta (u-r)$ as a function of
$q$ and $M_r^f$ implies for the properties of spiral galaxies
and the dust they contain.

\section{DATA}
\label{sec-data}

The Sloan Digital Sky Survey (SDSS) has imaged roughly
$\pi$ steradians of the sky \citep{yo00,ad08}. SDSS photometric
data is provided in five bands ($ugriz$) from the near ultraviolet
to the near infrared \citep{fu95,sm02}. The SDSS Data Release
6 (DR6) includes 9583 square degrees of photometric
coverage and 7425 square degrees of spectroscopic coverage
\citep{ad08}. The SDSS photometric data processing pipeline
performs a morphological star/galaxy separation, with
extended objects being labeled ``galaxies'' and point-like
objects being labeled ``stars''. For each galaxy, in each
photometric band, a pair of models are fitted to the two-dimensional
galaxy image. One model has a de Vaucouleurs profile \citep{dV48}:
\begin{equation}
I ( R ) = I_e \exp \left( - 7.67 [ (R / R_e)^{-1/4} - 1 ]
\right) \ ,
\end{equation}
truncated beyond $7 R_e$ to go smoothly to zero at $8 R_e$.
The other model has an exponential profile:
\begin{equation}
I (R) = I_e \exp \left[ - 1.68 ( R / R_e - 1) \right] \ ,
\end{equation}
truncated beyond $3 R_e$ to go smoothly to zero at $4 R_e$.
The SDSS DR6 pipeline also takes the best-fitting de Vaucouleurs
model and exponential model for each galaxy, and finds the
linear combination of the two that best fits the galaxy image.
The fraction of the total flux contributed by the de Vaucouleurs
component is the parameter fracDeV, which is constrained to
lie in the interval $0 \leq {\rm fracDeV} \leq 1$. The fracDeV
parameter thus acts as a concentration index, varying from
${\rm fracDeV} = 1$ for highly concentrated de Vaucouleurs
galaxies to ${\rm fracDeV} = 0$ for less concentrated exponential
galaxies.

As our measure of the flux of each galaxy, we use the
$r$ band ``model magnitude''; for galaxies with ${\rm fracDeV}
\geq 0.5$, the model magnitude is the integrated magnitude
of the de Vaucouleurs model, and for galaxies with
${\rm fracDeV} < 0.5$, it's the integrated magnitude of
the exponential model. As our measure of the color of
each galaxy, we use the $u-r$ color, defined as the
difference between the $u$ band model magnitude and the
$r$ band model magnitude. As our measure of the axis ratio
of each galaxy, we use the axis ratio of the $r$ band
25 mag arcsec$^{-2}$ isophote. The SDSS DR6 data pipeline
finds the best-fitting ellipse to the 25 mag arcsec$^{-2}$
isophote in each band; the semimajor axis and semiminor
axis of this isophotal ellipse are $A_{25}$ and $B_{25}$,
respectively. The isophotal axis ratio $q_{25} \equiv
B_{25}/A_{25}$ then provides a measure of the apparent
galaxy shape at a few times the effective radius; for
galaxies with ${\rm fracDeV} = 0$, the average value of
$A_{25} / R_e$ is about 2.4 \citep{vi05}.

Our full sample of galaxies consists of objects in
the SDSS DR6 spectroscopic sample that are labeled as
galaxies and that have spectroscopic redshifts
$z > 0.004$, to eliminate contaminating foreground
objects, and $z < 0.06$, to reduce the possibility
of weak lensing distortions of apparent shapes, and
to eliminate, in practice, the necessity of applying
$K$-corrections. To eliminate low-quality redshifts,
we require that the SDSS redshift confidence parameter
have a value ${\rm zConf} > 0.35$. To ensure that the
galaxies in our full sample are well resolved spatially,
we require that their photometric data fulfill the
criterion $\tau > 6.25 \tau_{\rm psf}$, where $\tau$
is the adaptive second-order moment of the galaxy image
and $\tau_{\rm psf}$ is the adaptive second-order moment
of the point spread function at the galaxy's location.
Our full galaxy sample, defined in this way, contains
$n = 78{,}230$ galaxies.

\begin{center}
\begin{figure}
\plotone{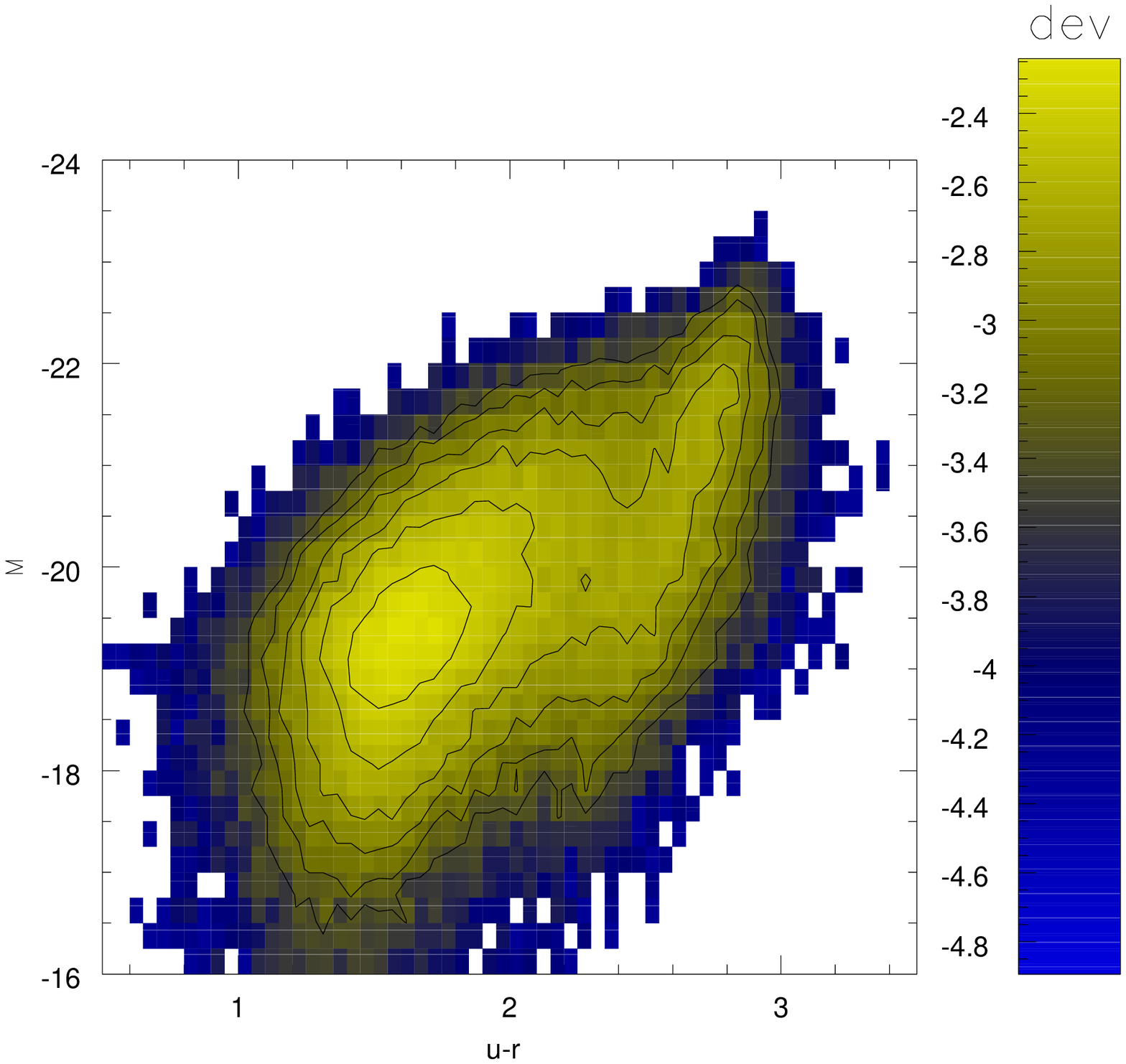}
\caption{Color -- magnitude diagram (absolute $r$ band magnitude versus
$u - r$) for all galaxies in our flux-limited $z < 0.06$ sample. The
color bar on the right indicates the value of $\log_{10} F$ corresponding
to each color, where $F$ is the fraction of the galaxies in each bin.
Bin sizes are $\Delta M_r = 0.25$ and $\Delta (u-r) = 0.05$.
}
\label{fig:1}
\end{figure}
\end{center}

Figure~\ref{fig:1} shows the color -- magnitude diagram
for the 78{,}230 galaxies in our full sample. The absolute
magnitude $M_r$ is computed from the model magnitude
$m_r$ assuming a Hubble constant $H_0 = 70 {\rm\,km}
{\rm\,s}^{-1}{\rm\,Mpc}^{-1}$ in a flat universe with
mass contributing $\Omega_{m,0} = 0.3$ to the density
parameter and a cosmological constant contributing
$\Omega_{\Lambda,0} = 0.7$. No $K$-corrections are
applied to the galaxies in our sample. In Figure~\ref{fig:1},
the division between the blue sequence, on the left,
and the red sequence, on the right, can be clearly seen.
However, the so-called ``green valley'', between the
blue and red sequences, is well populated with galaxies.

\begin{figure}
\plottwo{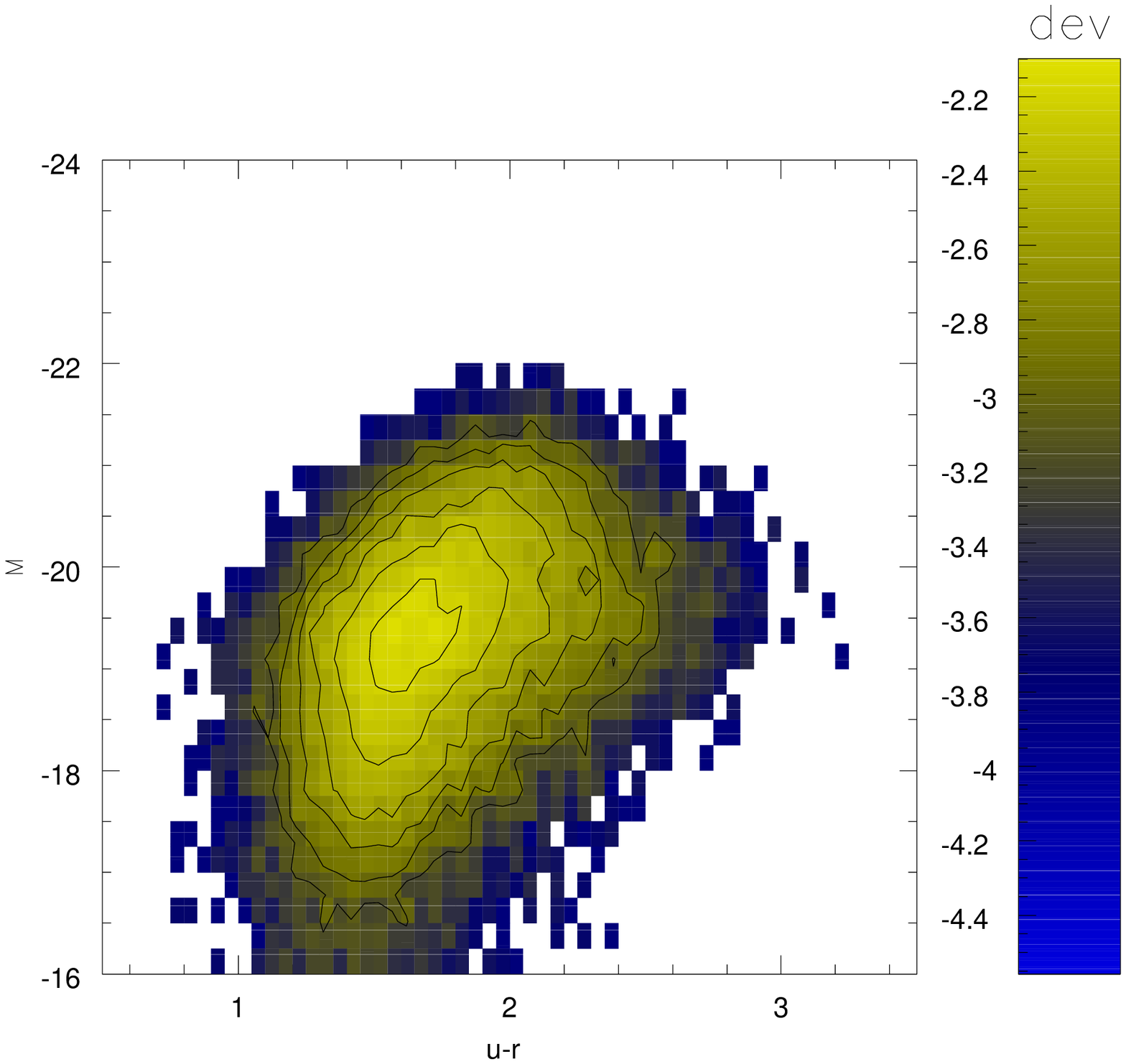}{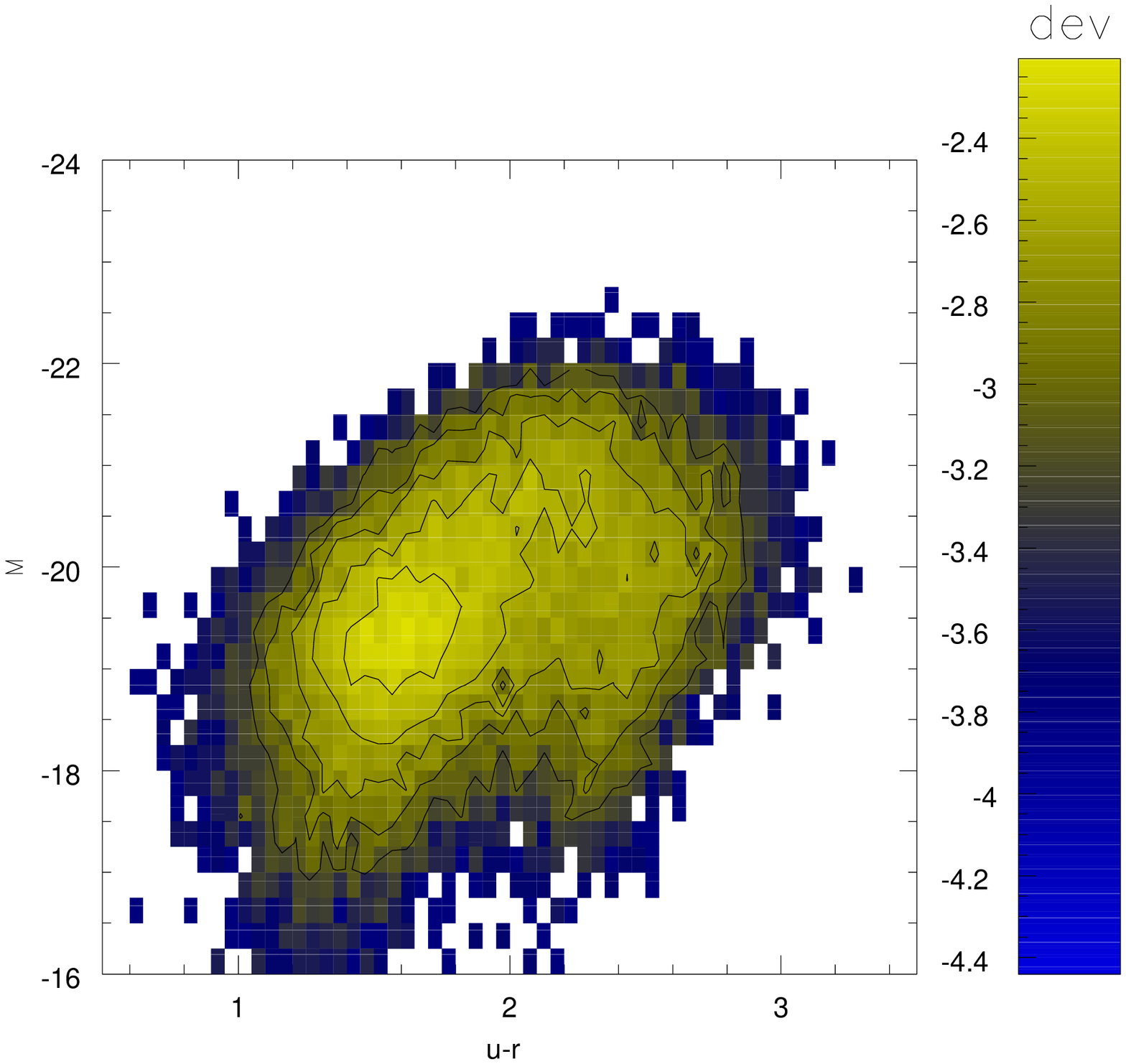}
\caption{Left: As in Figure~\ref{fig:1}, but containing only those
galaxies with ${\rm fracDeV} \leq 0.1$. Right: As in Figure~\ref{fig:1},
but containing only those galaxies with $0.1 < {\rm fracDeV} \leq 0.5$.
}
\label{fig:2}
\end{figure}

\begin{figure}
\plottwo{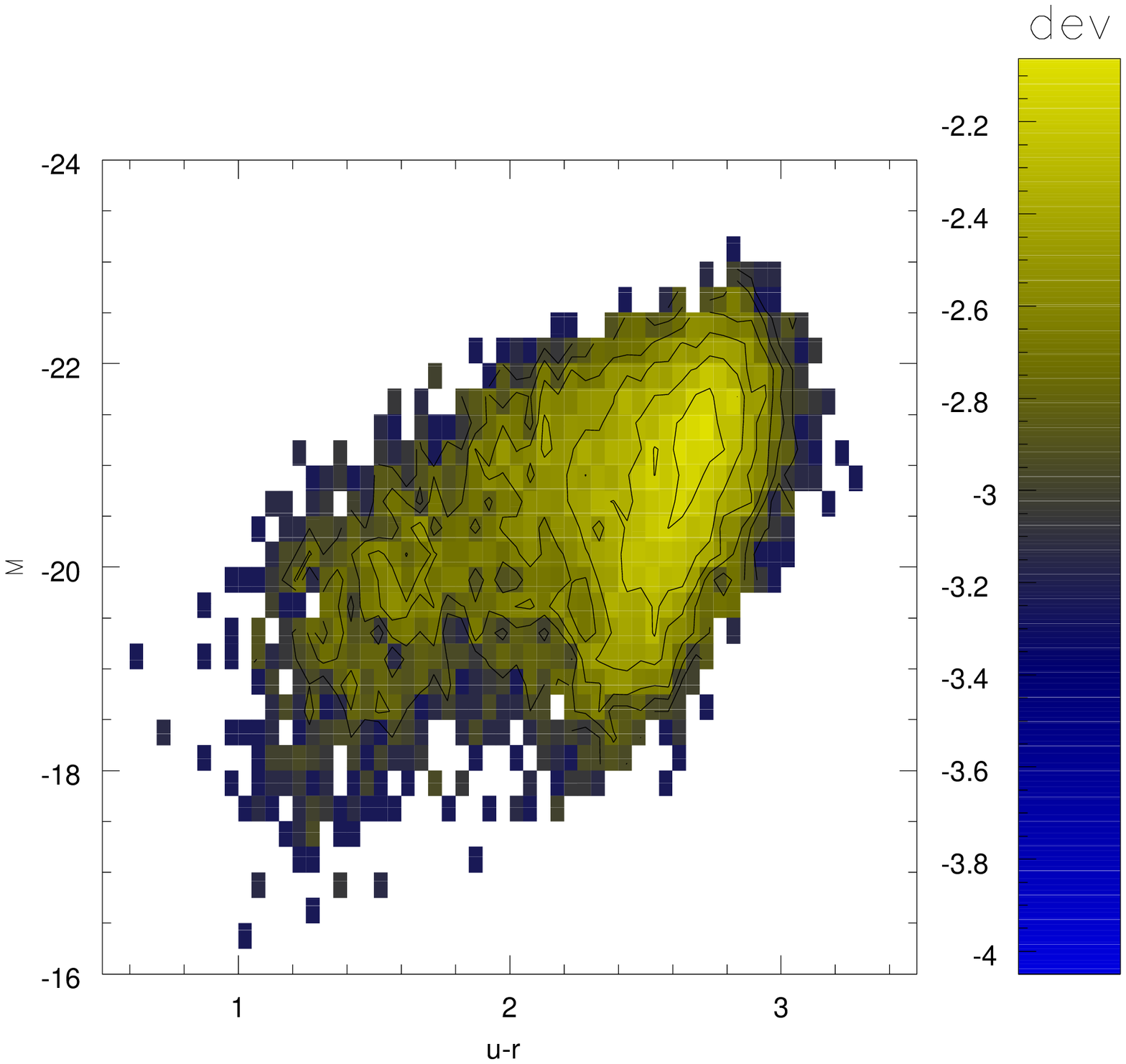}{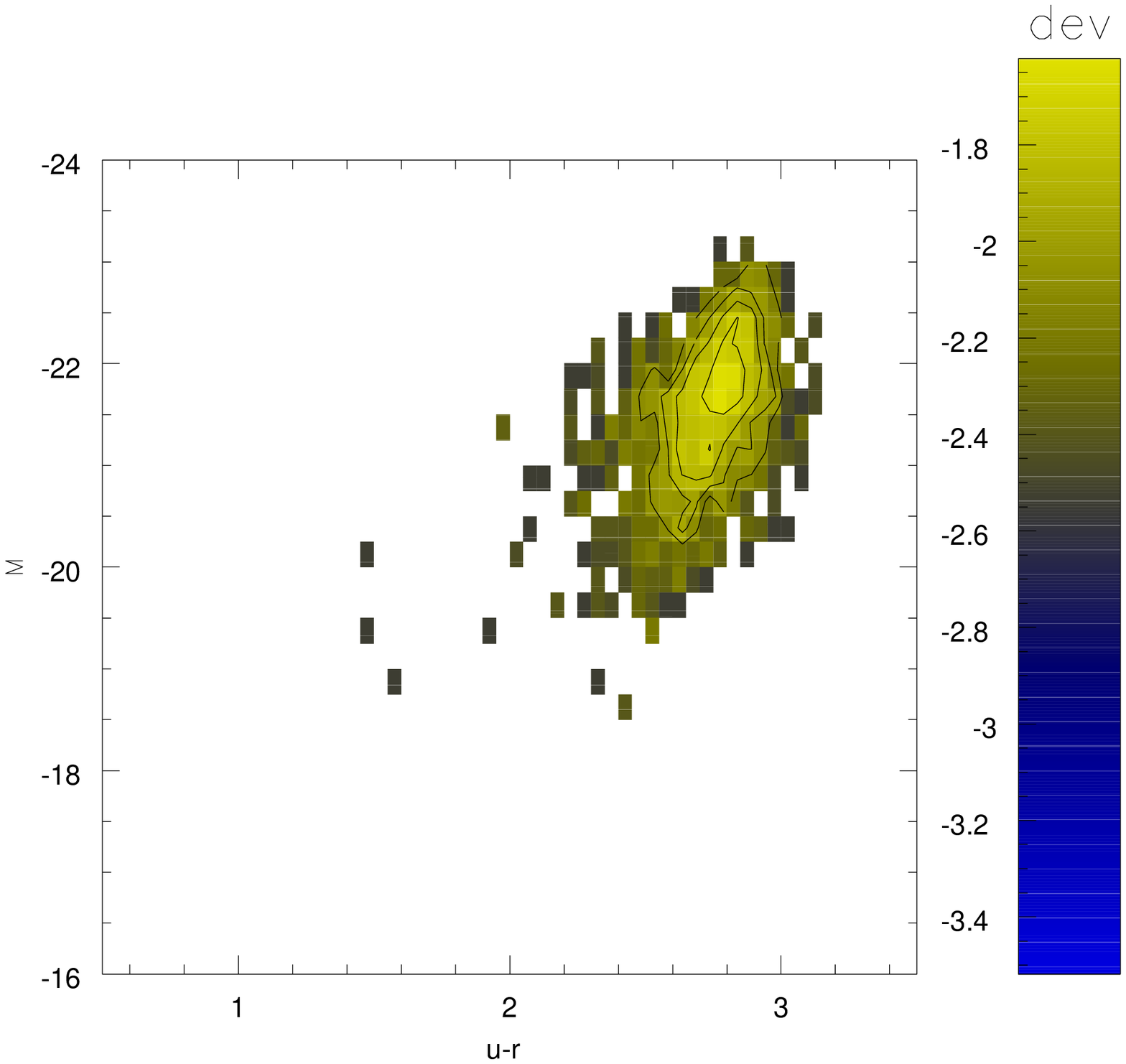}
\caption{Left: As in Figure~\ref{fig:1}, but containing only those
galaxies with $0.5 < {\rm fracDeV} \leq 0.9$. Right: As in Figure~\ref{fig:1},
but containing only those galaxies with ${\rm fracDeV} > 0.9$.
}
\label{fig:3}
\end{figure}

Using the information about surface brightness profiles
provided by the fracDeV parameter, we can isolate subsamples
that are all late-type or all early-type galaxies. Of
our full sample, $n = 36{,}162$ galaxies have
${\rm fracDeV} \leq 0.1$. The CM diagram for this exponential
subsample is shown in the left panel of Figure~\ref{fig:2}.
The exponential galaxies fall along a well-defined blue
sequence. Note, though, that the spread in $u-r$ color
increases with increasing luminosity. For comparison,
the right panel of Figure~\ref{fig:2} shows the CM
diagram for the $n = 27{,}567$ galaxies in our full
sample that have $0.1 < {\rm fracDeV} \leq 0.5$. These
more concentrated galaxies fall primarily along the
blue sequence; however, there are a significant number
at the faint end of the the red sequence
($M_r \sim -19.5$, $u-r \sim 2.4$). Moving to still
greater concentration, the left panel of Figure~\ref{fig:3}
shows the CM diagram for the $n = 11{,}202$ galaxies
that have $0.5 < {\rm fracDeV} \leq 0.9$. These
galaxies fall primarily along the red sequence; however,
there are a significant number at the bright end
of the blue sequence ($M_r \sim -20$, $u-r \sim 1.6$).
Finally, the right panel of Figure~\ref{fig:3} shows
the CM diagram for the relatively small number of
galaxies in the full sample ($n = 3299$) that have
${\rm fracDeV} > 0.9$. These very red, very luminous
de Vaucouleurs galaxies represent the high-luminosity
end of the red sequence.

Previous studies of the colors, absolute magnitudes, and
apparent shapes of late-type galaxies in the Sloan Digital
Sky Survey have used different definitions of ``late-type
galaxy''. \citet{bl03}, for instance, find that galaxies with
\citet{se68} index $n < 1.5$ fall on the blue sequence of a
CM diagram, while those with S\'ersic index $n > 3$ fall on the
red sequence. (A galaxy with a perfect S\'ersic profile will
have ${\rm fracDeV} \approx 0.3$ if $n = 1.5$ and ${\rm fracDeV}
\approx 0.8$ if $n = 3$.) \citet{ch06} and \citet{sh07}, in their
studies of late-type galaxies in the SDSS, find it useful to chose
a sample with ${\rm fracDeV} < 0.5$. This cut in fracDeV, however,
does allow some galaxies from the red sequence to enter the
sample. Since the fracDeV distribution is strongly peaked at
${\rm fracDeV} = 0$, we choose to make the more stringent
cut ${\rm fracDeV} \leq 0.1$ to create our late-type galaxy
subsample. The $n = 36{,}162$ galaxies with ${\rm fracDeV}
\leq 0.1$, we expect, will provide us with a sample of late-type
spirals whose light is dominated by a dusty exponential disk.

\section{ANALYSIS}
\label{sec-analysis}

\begin{figure}
\plotone{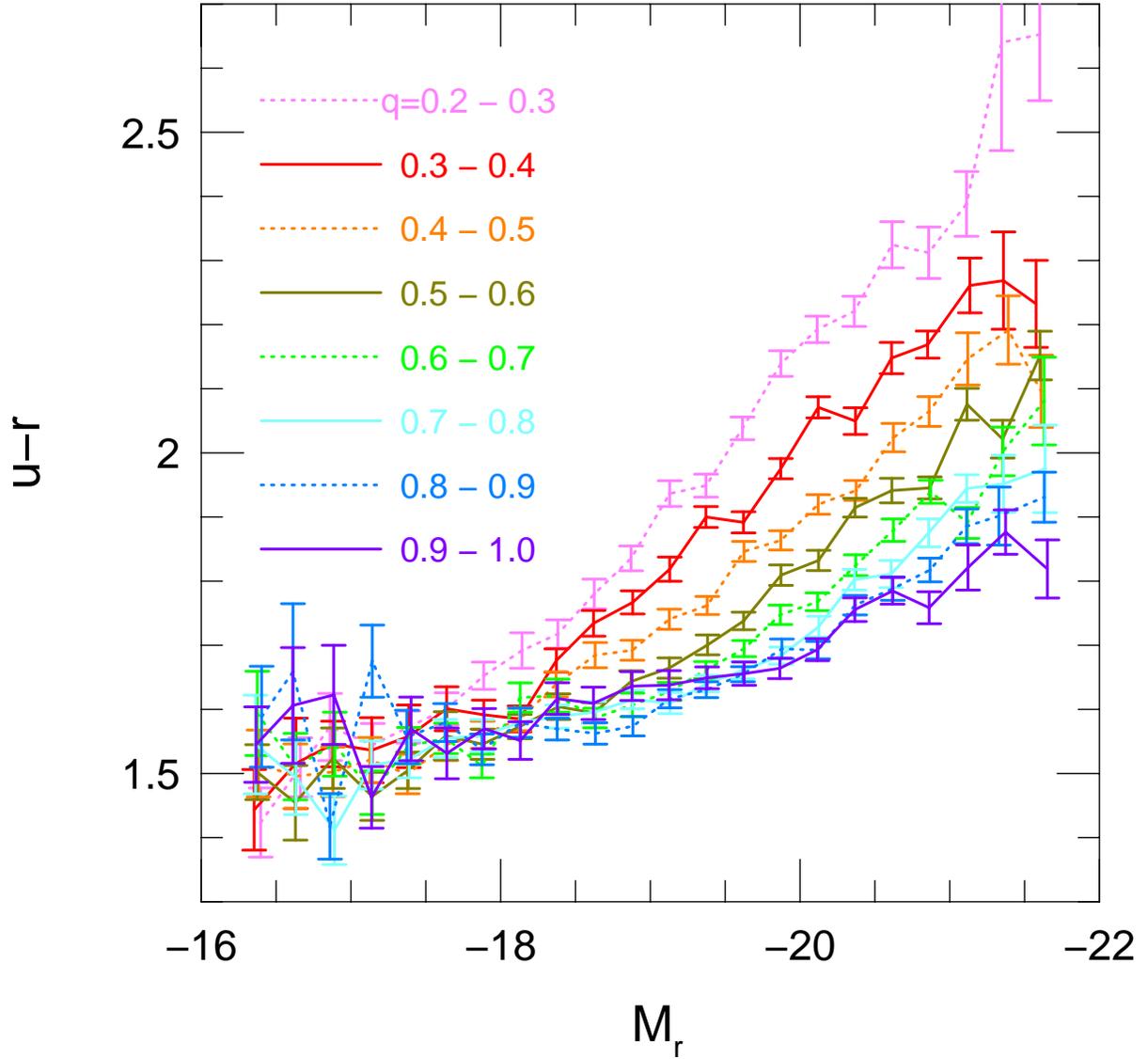}
\caption{Average $u-r$ color as a function of $M_r$ for our flux-limited,
$z < 0.06$ sample of exponential galaxies. Results are shown for different ranges
of $q$, the apparent axis ratio. Error bars represent the estimated error in the
mean color.
}
\label{fig:4}
\end{figure}

How do we know that our subsample of (nearly) exponential
galaxies are actually flattened disks? We have, after all,
chosen them solely on the basis of their surface brightness
profiles. Although the disks of spiral galaxies are known
to be well fitted by exponential profiles, so are other
subclasses of galaxies, such as dwarf ellipticals. We can
test the assertion that exponential galaxies are flattened,
dust-containing disks by looking at their colors as a function
of apparent axis ratio. If a galaxy is a disk, then its apparent
axis ratio $q$ will be smallest when it is edge-on. If the galaxy's
disk is dusty, then the galaxy will be most reddened when it
is edge-on. Figure~\ref{fig:4} is a plot of the mean $u-r$
color as a function of absolute magnitude $M_r$ for the
$n = 36{,}162$ galaxies in our exponential subsample
(${\rm fracDeV} \leq 0.1$). The galaxies are binned by
their apparent axis ratio $q$. Note that for exponential
galaxies brighter than $M_r \sim -19$, there is a noticeable
correlation between $q$ and $u-r$ at a given absolute magnitude,
with smaller values of $q$ corresponding to larger values of
$u-r$. This is just what we expect for a population of dusty
disk galaxies. However, at $M_r > -18$, there is no correlation
between $q$ and $u-r$. At these low luminosities, the galaxies
in the exponential subsample are blue ($u-r \sim 1.5$) dwarf
galaxies in which the stars and dust are not in orderly thin disks.
Thus, in looking for inclination-dependent colors and magnitudes,
we will look at those high-luminosity ($M_r < -19$) exponential
galaxies for which the color actually is inclination-dependent.

In general, when a spiral galaxy is seen at an arbitrary inclination,
it will be fainter at visible wavelengths than it would be
seen face-on. We may write
\begin{equation}
M_r = M_r^f + \Delta M_r \ ,
\end{equation}
where $M_r$ is the $r$-band absolute magnitude computed from
the actual apparent magnitude and redshift, $M_r^f$ is what the
absolute magnitude would be if the galaxy were seen face-on, and
$\Delta M_r \geq 0$ is the inclination-related dimming. In
general, $\Delta M_r$ will be a function of the (unknown)
inclination $i$ as well as of the detailed properties of
the observed galaxy's dust. We will assume that the apparent
axis ratio $q$ of the 25 mag arcsec$^{-2}$ isophote will be
an adequate measure of the inclination. If every spiral galaxy
were a perfect oblate spheroid, with intrinsic short-to-long
axis ratio $\gamma = c/a$, then the inclination $i$ would be
uniquely determined by the apparent axis ratio $q$, through the
usual relation 
\begin{equation}
\cos^2 i = ( q^2 - \gamma^2 ) / ( 1 - \gamma^2 ) \ .
\label{eq:inc}
\end{equation}
One source of error in this relation is that spiral galaxies
don't all have the same intrinsic thickness $\gamma$. However,
an erroneous assumed value for $\gamma$ is irrelevant in the case
of low inclination and high $q$: a face-on thick disk has the same
apparent shape as a face-on thin disk. An erroneous assumed
value for $\gamma$ produces the largest error in $i$ when
the apparent axis ratio $q$ is small: a circular disk with
$q = 0.3$ is at an inclination $\cos i = 0$ if it's
a fat disk with $\gamma = 0.3$, but at an inclination
$\cos i = 0.28$ if it's an ultrathin disk with $\gamma = 0.1$.
Another source of error in equation~(\ref{eq:inc}) is that
disks are not perfectly circular. Consider an ultrathin
disk with $\epsilon \approx 0.1$, a typical ellipticity for a
spiral galaxy. When the galaxy is face-on, equation~(\ref{eq:inc})
will yield $\cos i \approx 0.9$, instead of the true value
of $\cos i = 1$. When the same disk is viewed at a high
inclination, the value of $\cos i$ we compute will
have an error $\sim \epsilon \cos 2 \phi$, where
$\phi$ is the azimuthal viewing angle measured
relative to the intrinsic long axis of the disk
\citep{ry06}. Thus, when we use $q$ as a surrogate for the
inclination $i$, we should remember that two spiral
galaxies with the same $q$ may have $\cos i$ differing
by as much as $\sim \epsilon$, where $\epsilon$ is their
average intrinsic disk ellipticity.

To simplify our analysis, we will start by assuming that
$\Delta M_r$ is a function only of $q$, and not of $M_r^f$.
That is, we assume that all spiral galaxies with a given
$q$ suffer the same fractional loss of flux in the $r$ band.
This is not necessarily correct -- bright spirals may well
have systematically different dust properties from dimmer
spirals, for instance -- but it provides a place to start.
To find $\Delta M_r (q)$, we begin by creating a volume-limited
subsample of our exponential (${\rm fracDeV} \leq 0.1$)
SDSS galaxies. The redshift limit of the volume-limited
subsample is $z = 0.06$; the flux limit is assumed to be
$m_r = 17.77$, resulting in a low-luminosity cutoff of
$M_r = -19.4$. (The SDSS spectroscopic survey is complete
to a limiting Petrosian apparent magnitude $r = 17.7$; for
galaxies that are well fitted by the exponential model galaxy,
the Petrosian apparent magnitudes and the model magnitudes
used in our analysis have an rms scatter of $< 0.1 {\rm\,mag}$.)
We then take the $n = 16{,}363$ exponential galaxies in
the volume-limited subsample and bin them by apparent axis
ratio, in bins of width $\Delta q = 0.1$. The galaxies with
$0.9 < q \leq 1.0$ are galaxies that we expect to be nearly
face-on. 

\begin{figure}
\plotone{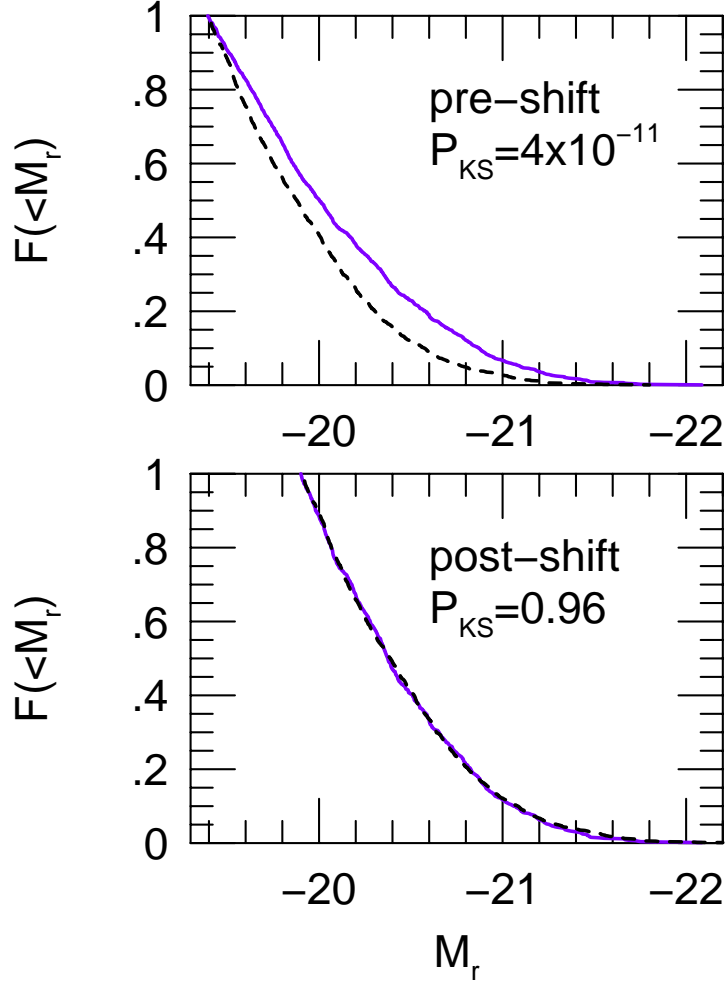}
\caption{Top: Cumulative luminosity function for the galaxies in
the volume-limited sample with $q > 0.9$ (solid line) and with
$0.2 < q \leq 0.3$ (dotted line). Functions are normalized so
that $F ( < M_r ) = 1$ at $M_r = -19.40$.
Bottom: Comparison of the cumulative luminosity functions after
the $0.2 < q \leq 0.3$ subsample has been shifted by $-\Delta M_r =
-0.51$. Functions are normalized so that
$F ( < M_r ) = 1$ at $M_r = -19.40 - 0.51$.
}
\label{fig:5}
\end{figure}

The cumulative luminosity function for the exponential
galaxies with $0.9 < q \leq 1.0$ is shown as the solid line
in the upper panel of Figure~\ref{fig:5}. Since we expect
$\Delta M_r = 0$ for these nearly face-on galaxies, this solid
line is our estimator of the normalized luminosity function
$f (M_r^f)$ for face-on spiral galaxies brighter than
$M_r^f = -19.4$. For comparison, the dashed line in the upper
panel of Figure~\ref{fig:5} shows the cumulative luminosity
function for exponential galaxies with $0.2 < q \leq 0.3$;
we expect these galaxies to nearly edge-on. The luminosity
function for the $q \sim 0.25$ galaxies is different from
that of the $q \sim 0.95$ galaxies; a Kolmogorov-Smirnov
test yields a probability $P_{\rm KS} = 4 \times 10^{-11}$.
This difference results from the fact that the luminosity
function of spiral galaxies has a cutoff at high luminosity;
for edge-on galaxies, this cutoff is dimmed by an amount
$\Delta M_r$. Using our assumption that dimming (in magnitudes)
is independent of $M_r^f$, we take every galaxy in the
$0.2 < q \leq 0.3$ and brighten it by the same amount $\Delta M_r$.
We then compare this artificially brightened luminosity
function to the luminosity function for face-on ($0.9 < q \leq 1.0$)
galaxies brighter than $M_r = -19.4-\Delta M_r$. The
comparison is done between the cumulative luminosity functions
using a Kolmogorov-Smirnov test. The bottom panel of
Figure~\ref{fig:5} shows the comparison between the $0.2 < q \leq 0.3$
galaxies and the $0.9 < q \leq 1.0$ galaxies, using the
optimal shift $\Delta M_r = 0.51$. The Kolmogorov-Smirnov
test yields a probability $P_{\rm KS} = 0.96$ for this shift;
our hypothesis of $\Delta M_r$ independent of $M_r^f$ cannot
be excluded, at least in the luminosity range $M_r^f < -19.91$.

\begin{figure}
\plotone{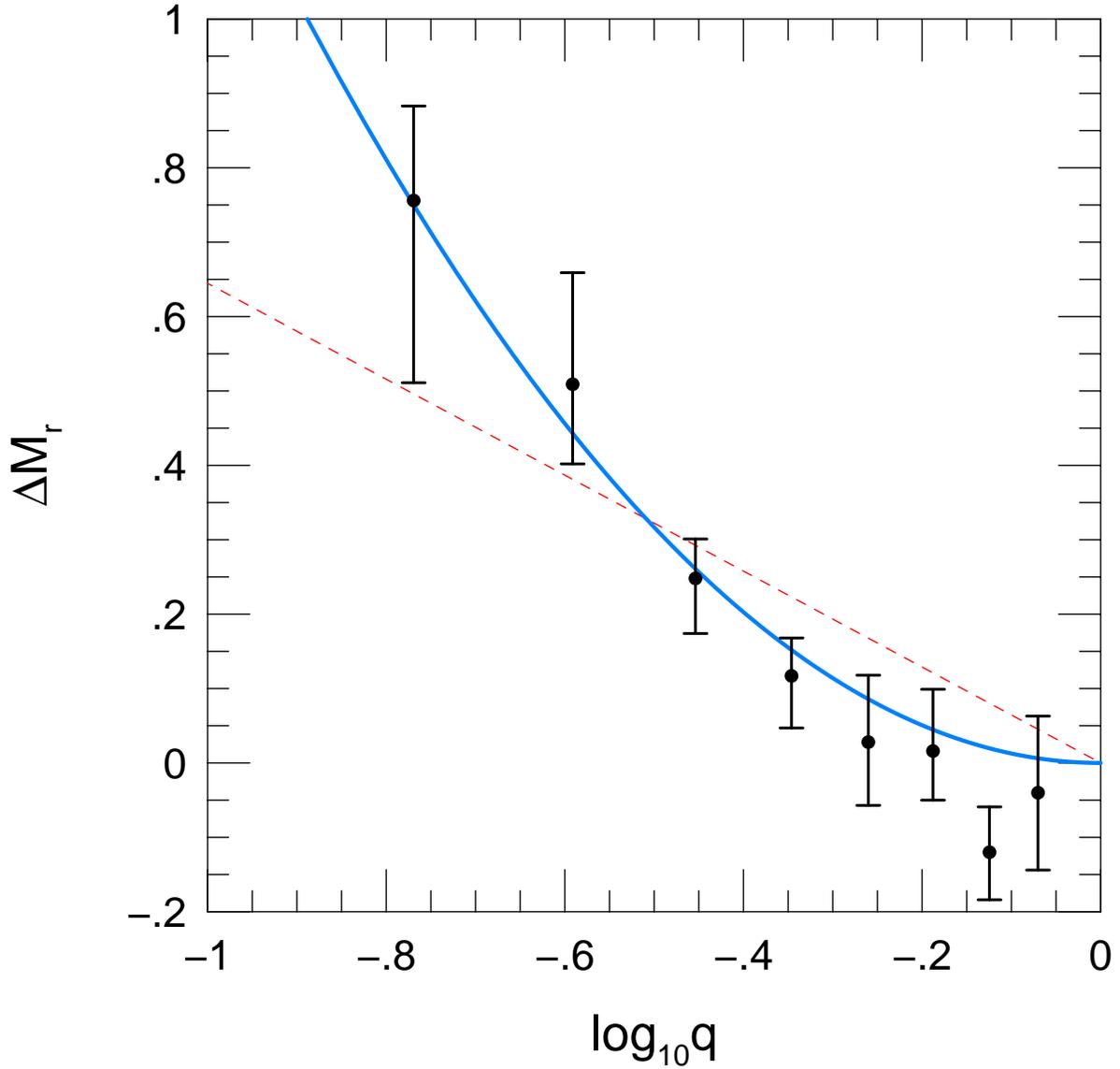}
\caption{The points show the best-fitting shift $\Delta M_r$ of
the exponential galaxy luminosity function as a function of
the apparent axis ratio $q$. The error bars indicate the
range of $\Delta M_r$ for which $P_{\rm KS} > 0.1$. The
dashed red curve shows $\Delta M_r = -0.64 \log q$,
and the solid blue curve shows $\Delta M_r = 1.27
( \log q )^2$.
}
\label{fig:6}
\end{figure}

Figure~\ref{fig:6} gives a summary of the best-fitting absolute magnitude
shift $\Delta M_r$ as a function of $q$ for the exponential galaxies.
Each data point gives the value of $\Delta M_r$ which maximizes
the Kolmogorov-Smirnov probability when the luminosity function of
galaxies in a particular range of $q$ are compared to galaxies with
$q > 0.9$. The error bars indicate the range of $\Delta M_r$ over
which $P_{\rm KS} > 0.1$. Because the high-luminosity cutoff in
the luminosity function is not an extremely sharp feature, the
range of statistically acceptable values of $\Delta M_r$ is as
large as $\sim 0.4 {\rm\,mag}$ in the low-$q$ bins, where there
are relatively few galaxies. A common parameterization of the
$\Delta M$ -- $q$ relation, following the Third Reference
Catalogue \citep{dV91}, is
\begin{equation}
\Delta M = - \gamma_\lambda \log q \ ,
\label{eq:logq}
\end{equation}
where the value of $\gamma_\lambda$ depends on the wavelength of observation,
and on the population of galaxies observed. \citet{dV91} used
$\gamma_B = 1.5$ for spiral galaxies of type Sc observed in the
$B$ band. For pure disk systems (type Sc-Sd), \citet{bo95} found
$\gamma_B = 1.67$. \citet{tu98} found dimming that was dependent
on luminosity as well as on wavelength: for bright spirals in
the Ursa Major and Pisces clusters, they found $\gamma_B = 1.7$,
$\gamma_R = 1.3$, and $\gamma_I = 1.0$ in the $B$, $R$, and $I$
bands. \citet{sh07}, looking at ${\rm fracDeV} < 0.5$ galaxies
in the SDSS, found values of $\gamma$ ranging from $\gamma_u = 2.19$
in the $u$-band, through $\gamma_r = 1.37$ in the $r$-band, to
$\gamma_z = 0.80$ in the $z$-band.

Our best fit to the logarithmic relation of equation~(\ref{eq:logq})
is shown as the dashed line in Figure~\ref{fig:6}; the best fitting
value of $\gamma$ is $\gamma_r = 0.64$. Note, however, that the
$\Delta M_r \propto \log q$ fit is not very good. It overestimates
the dimming of galaxies with $q > 0.4$, and underestimates the
dimming of nearly edge-on galaxies with $q < 0.3$. In fact, our
data are consistent with no dimming at all in the $r$ band
for galaxies with $q > 0.5$. A superior fit is provided by the
solid line in Figure~\ref{fig:6}, which represents a dimming
proportional to the square of $\log q$:
\begin{equation}
\Delta M_r = \beta_r ( \log q )^2 \ ,
\label{eq:log2q}
\end{equation}
with $\beta_r = 1.27$.
Our results are in qualitative agreement with those of \citet{ma03},
who find that for galaxies in the Two Micron All-Sky Survey
Extended Source Catalog, the simple linear extinction law of
equation~(\ref{eq:logq}) gives an inadequate fit to the
dimming in near-infrared ($J$, $H$, and $K_s$) bands. A
better fit is provided by a bilinear function with a steeper
slope at $\log q < -0.5$ than at $\log q > 0.5$. A good fit
is also provided by a quadratic in $\log q$. In the $J$ band,
for instance, \citet{ma03} find the best quadratic fit is
$\Delta M_J = 0.12 \log q + 1.14 ( \log q )^2$.
If we attempt a fit of this form to our exponential
$r$-band subsample, we find
$\Delta M_r = 0.25 \log q + 1.66 ( \log q )^2$.
However, this does not provide a statistically better fit,
given the loss of a degree of freedom, than the simpler
parabolic form of equation~(\ref{eq:log2q}).

\begin{figure}
\plotone{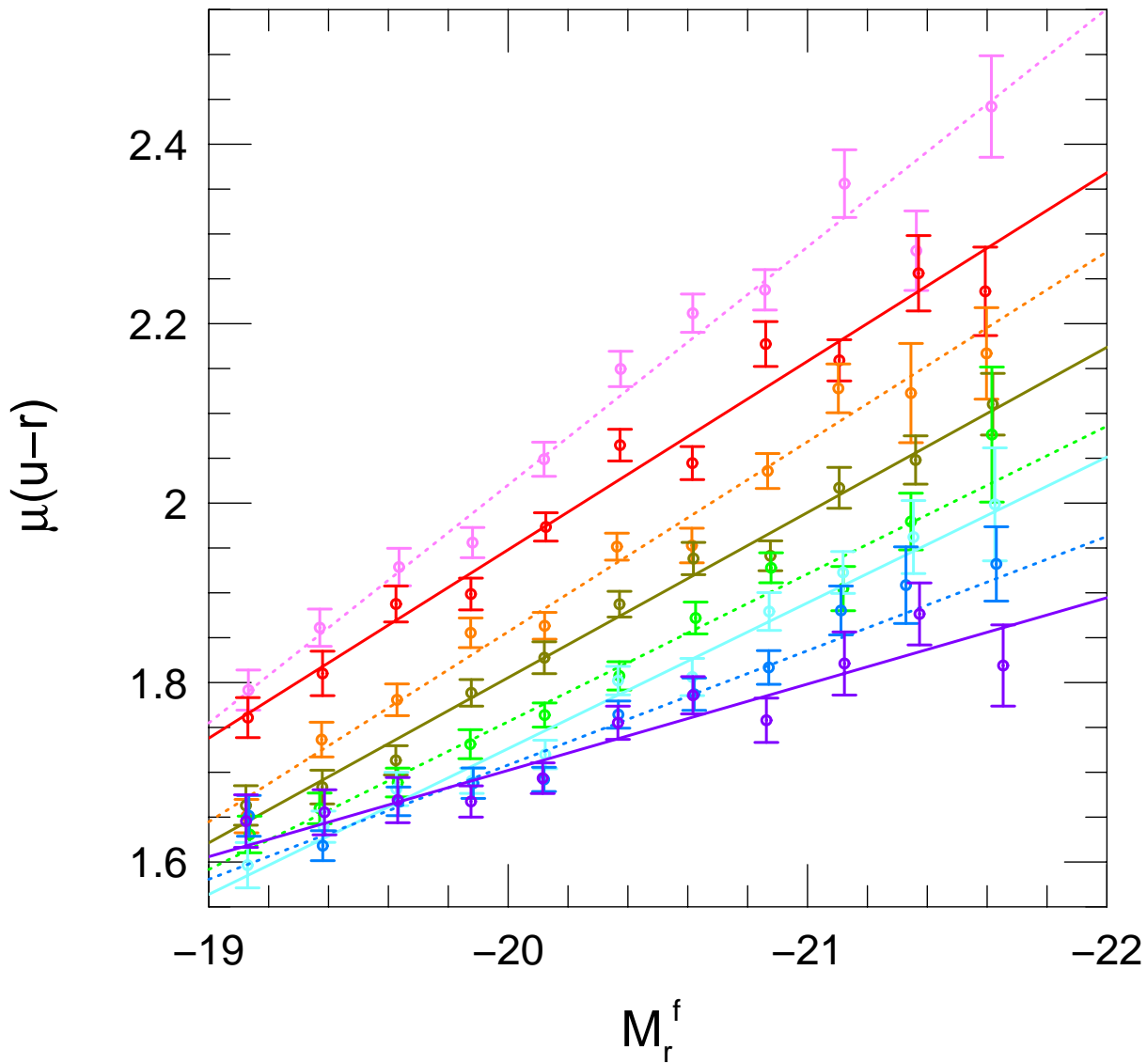}
\caption{Average $u-r$ color as a function of the corrected,
``face-on'' absolute magnitude $M_r^f$ for our flux-limited,
$z < 0.06$ sample of exponential galaxies. Colors and line
types are the same as in Figure~\ref{fig:4}. Error bars represent
the estimated error in the mean color.
}
\label{fig:7}
\end{figure}

Given the dimming correction of equation~(\ref{eq:log2q}), we
can compute the ``face-on'' absolute magnitude $M_r^f$ for
every exponential galaxy with measured $m_r$, $z$, and $q$:
\begin{equation}
M_r^f = M_r - \Delta M_r = M_r - 1.27 ( \log q )^2 \ .
\label{eq:mrf}
\end{equation}
Figure~\ref{fig:7} shows the average $u-r$ color for
exponential galaxies, as a function of the corrected
absolute magnitude $M_r^f$. We look only at galaxies
with $M_r^f < -19$, to exclude the dwarf galaxies for
which the correction is inappropriate. The different
colors and line types in Figure~\ref{fig:7} represent
different values of the apparent axis ratio $q$, just as
in Figure~\ref{fig:4}. For each range of $q$, the mean
color, $\langle u-r \rangle$, is linear in the corrected
absolute magnitude, $M_r^f$, with brighter galaxies being redder,
on average. The $q > 0.9$ galaxies, which are nearly face-on,
have a relatively small dependence of average color on
absolute magnitude; for the $q > 0.9$ galaxies, increasing the
brightness by $1 {\rm\,mag}$ in the $r$ band corresponds to
reddening the galaxy by $\sim 0.1 {\rm\,mag}$, on average, in $u-r$.
This correlation is a manifestation of the dependence of
stellar population on galaxy luminosity; in general, more
luminous spiral galaxies have populations that are older
and more metal-rich \citep{be00,ma04}. The nearly edge-on
galaxies ($q < 0.3$), have a greater dependence of average color
on absolute magnitude; for the $q < 0.3$ galaxies, an increase
in brightness by $1 {\rm\,mag}$ in the $r$ band corresponds
to a reddening of $\sim 0.27 {\rm\,mag}$, on average, in $u-r$.
The steepness of the mean color -- luminosity relation for
nearly edge-on spiral galaxies is a manifestation of the
dependence of dust opacity on absolute magnitude; in general,
more luminous spiral galaxies, being more metal-rich, have
greater disk opacity due to dust \citep{ma03}.

\begin{figure}
\plottwo{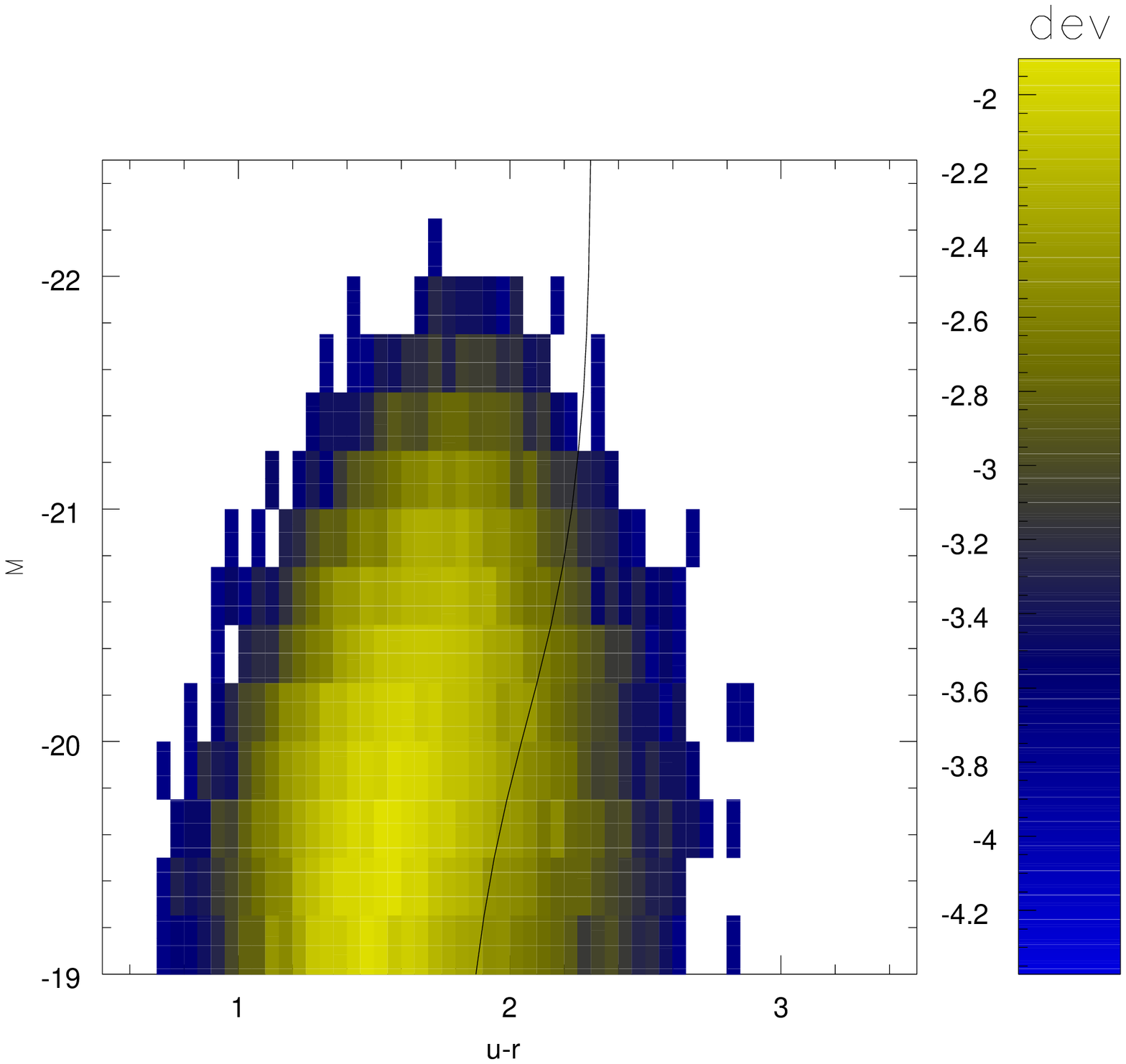}{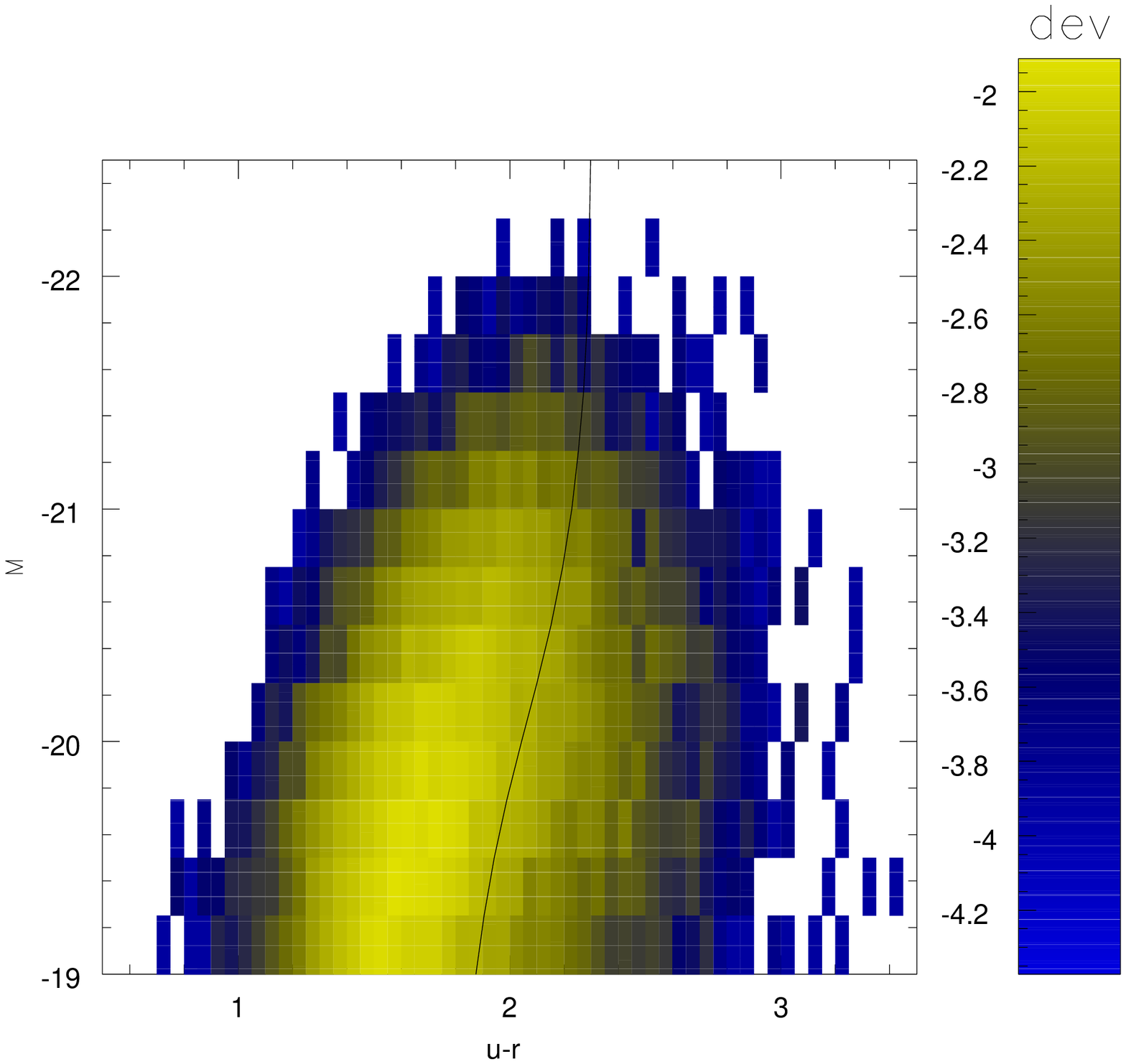}
\caption{Left: Color -- magnitude diagram for ${\rm fracDeV} < 0.1$
galaxies, using the corrected $(u-r)^f$ color (equation~\ref{eq:urf})
and the corrected $M_r^f$ absolute magnitude (equation~\ref{eq:mrf}).
Right: Same as left panel, but using uncorrected $u-r$ color and
$M_r$ absolute magnitude.
}
\label{fig:8}
\end{figure}

By doing a least-squares fit to the function
\begin{equation}
\langle u-r \rangle = a + b ( M_r^f + 20.5 ) \ ,
\end{equation}
we find the straight lines in Figure~\ref{fig:7}; the
intercepts ($a$) and slopes ($b$) are given in
Table~\ref{tab:1}. The mean $u-r$ color for an exponential
galaxy with $M_r^f = -20.5$ is nearly linear in $\log q$;
our best fit is
\begin{equation}
a = 1.72 - 0.723 \log q \ .
\end{equation}
The best fitting function of a similar form for $b$ as a function
of $q$ is
\begin{equation}
b = -0.11 + 0.255 \log q \ .
\end{equation}
We can compute a corrected, ``face-on'' color, $(u-r)^f$, for
each galaxy using the relation
\begin{equation}
(u-r)^f = (u-r) + [ 0.723 - 0.255 (M_r^f + 20.5 ) ] \log q \ ,
\label{eq:urf}
\end{equation}
where $u-r$ is the color computed from observations, and
$M_r^f$ is the corrected absolute magnitude from
equation~(\ref{eq:mrf}). To illustrate the effects of
using the ``face-on'' colors and absolute magnitudes on
the CM diagram, Figure~\ref{fig:8} shows the CM diagram
for the exponential (${\rm fracDeV} < 0.1$) galaxies in
our flux-limited $z < 0.06$ sample. The left panel uses
the corrected absolute magnitudes and colors from
equations~(\ref{eq:mrf}) and (\ref{eq:urf}), while the
right panel uses the uncorrected values. Of the
21{,}813 galaxies with $M_r < -19$ plotted in the
right panel of Figure~\ref{fig:8}, 5213 (23.9\%) have
a $u-r$ color redder than the optimal divider of
\citet{ba04}, intended to divide the red sequence
from the blue sequence with optimal effectiveness.
By contrast, of the  23{,}602 galaxies with $M_r^f < -19$
plotted in the left panel, only 2335 (9.9\%) have colors
redder than the optimal dividing color of \cite{ba04}.

\begin{figure}
\plotone{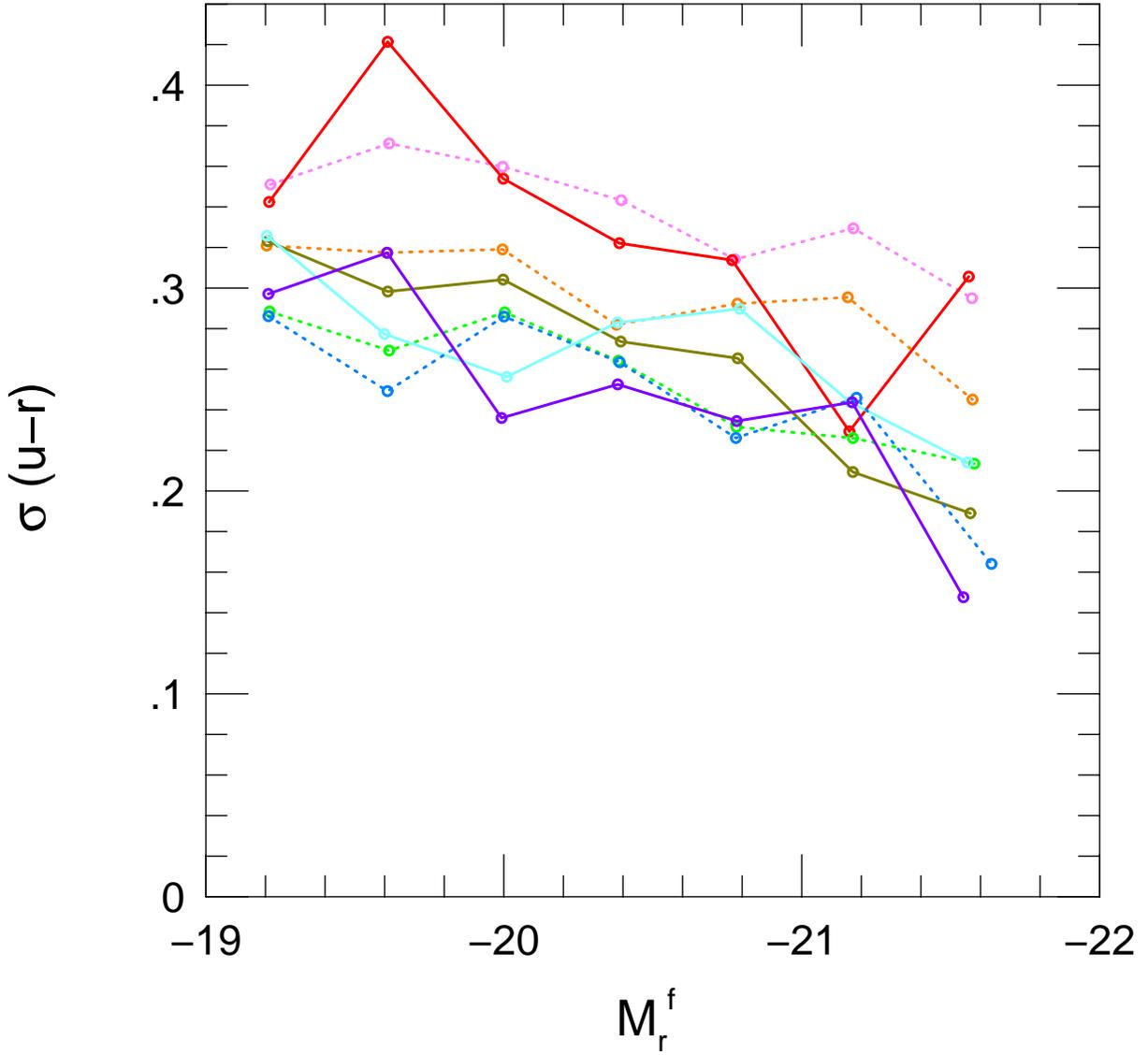}
\caption{Standard deviation in the $u-r$ color as a function of
the corrected absolute magnitude $M_r^f$ for our flux-limited,
$z < 0.06$ sample of exponential galaxies. Colors and line
types are the same as in Figure~\ref{fig:4}.
}
\label{fig:9}
\end{figure}

The color correction of equation~(\ref{eq:urf}) is
an average correction; the actual correction for any
individual galaxy will be different, due to the
galaxy-to-galaxy variation in dust properties. The
consequence of the variation is seen in Figure~\ref{fig:9},
which shows the standard deviation in the $u-r$ color,
as a function of $M_r^f$, for different values of $q$.
For nearly face-on ($q > 0.9$) exponential galaxies,
the standard deviation in color is luminosity dependent,
decreasing from $\sigma (u-r) \sim 0.3$ at $M_r^f \sim -19$
to $\sigma (u-r) \sim 0.2$ at $M_r^f \sim -21.5$. For
nearly edge-on galaxies ($q < 0.3$), the standard deviation
is greater at any given value of $M_r^f$, ranging from
$\sigma (u-r) \sim 0.35$ at $M_r^f \sim -19$ to
$\sigma (u-r) \sim 0.3$ at $M_r^f \sim -21.5$. At any
value of $M_r^f$, the edge-on galaxies have an excess
in $\sigma (u-r)$, caused by variations in dust properties,
of $\Delta \sigma \sim 0.23$, added in quadrature.

\section{INTRINSIC SHAPES}
\label{sec-shapes}

Our analysis has implicitly assumed that bright
SDSS galaxies with ${\rm fracDeV} < 0.1$ are
flattened, nearly circular disks. Now that we
have an empirical correction for dimming as a
function of apparent axis ratio (equation~\ref{eq:mrf}),
we can test whether this assumption is self-consistent.
If a population of galaxies consists of oblate spheroids
with a random distribution of inclinations, the observed
distribution of apparent axis ratios, $f(q)$, can be
inverted to find the distribution of intrinsic axis
ratios, $f (\gamma)$ \citep{hu26,sa70}.

For a population of dusty disk galaxies whose apparent
magnitude depends on inclination, selecting a subsample
whose inclinations are random requires a little extra care.
A simple flux-limited subsample will be biased against
disks at high inclination, which will have a lower flux than
a low-inclination disk at the same redshift with the same
luminosity. Thus, we create a corrected flux-limited sample;
for each SDSS exponential (${\rm fracDeV} < 0.1$) galaxy,
we ask, not simply whether it is above our flux limit, but
whether it would be above our flux limit if it were edge-on.
If an SDSS exponential galaxy has $q = 0.2$, we assume that
it is already edge-on. If it has an observed apparent axis
ratio $q_{\rm obs} > 0.2$, we compute its edge-on flux to be
\begin{eqnarray}
m_r ({\rm edge-on}) &=& m_r ({\rm observed}) + \Delta M_r (q = 0.2)
- \Delta M_r (q_{\rm obs}) \\
&=& m_r ({\rm observed}) + 1.27 [
(\log 0.2)^2 - ( \log q_{\rm obs} )^2 ] \ .
\end{eqnarray}
We create our corrected flux-limited sample by demanding that
this computed edge-on flux be greater than $m_r = 17.77$, the
completeness limit of the SDSS spectroscopic survey for galaxies.
To ensure that our inclination-corrected sample contains only the
luminous, nearby galaxies for which our $\Delta M_r (q)$ correction
was computed, we add the addition restrictions $M_r^f < -19.4$
and $z < 0.06$.

\begin{figure}
\plotone{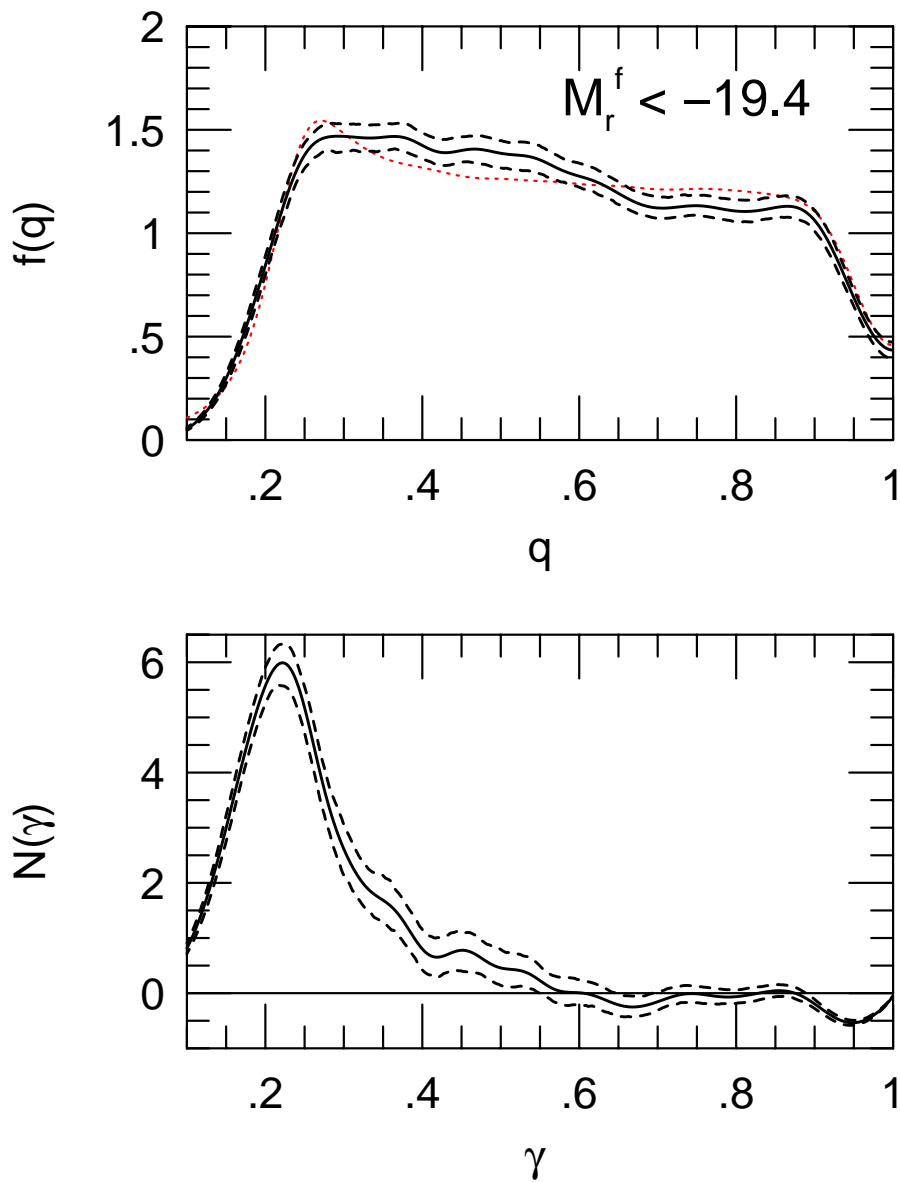}
\caption{Top: Distribution of apparent axis ratios for the
inclination-corrected sample of SDSS exponential galaxies.
The solid line is the best fit: the dashed lines show the
98\% confidence interval estimated from bootstrap resampling. The
red dotted line is the distribution of $q$ yielded by the best-fitting
parametric distribution of elliptical disks (eqns.~\ref{eq:gamma} and
\ref{eq:epsilon}).
Bottom: Distribution of intrinsic axis ratios, assuming galaxies are
randomly inclined oblate spheroids. The solid line is the inversion
of the best fit for $f(q)$; the dashed lines are the 98\% confidence
interval from bootstrap resampling.
}
\label{fig:10}
\end{figure}

The distribution of the apparent axis ratio $q$ for the $n = 16{,}155$
exponential galaxies in our inclination-corrected sample is shown
in the top panel of Figure~\ref{fig:10}. The distribution shown
as the solid line is estimated using a nonparametric kernel
technique \citep{vi94,tr95}. To ensure that our estimate of
$f(q)$ is smooth, we use a Gaussian kernel. The kernel width
is chosen using the standard adaptive two-stage estimator of
\citet{ab82}. To ensure that our estimate of $f(q)$ is zero for
$q < 0$ and $q > 1$, we impose reflective boundary conditions
at $q = 0$ and $q = 1$. The dashed lines in Figure~\ref{fig:10}
indicate the 98\% error intervals on $f(q)$ found by bootstrap
resampling. In our bootstrap analysis, we randomly selected
$n = 16{,}155$ values of $q$, with substitution, from the
original set of $n$ data points, and then created a new estimate
of $f(q)$ for the bootstrapped data. After doing 500 such bootstrap
estimate, we then determined the 98\% error intervals show in
Figure~\ref{fig:10}. At every value of $q$, 1\% of the bootstrap
estimates lie above the upper dashed line, and 1\% lie below the
lower dashed line.

Having a smooth estimate of $f(q)$, the distribution of apparent
axis ratios, allows us to compute $N(\gamma)$, the distribution of
intrinsic axis ratios, given the assumption that all galaxies are
oblate or prolate. If the disk galaxies are assumed to be oblate,
the relation between $f(q)$ and $N(\gamma)$ is
\begin{equation}
f (q) = \int_0^q P_{\rm obl} ( q | \gamma ) N ( \gamma ) d \gamma \ ,
\label{eq:volt}
\end{equation}
where $P_{\rm obl} ( q | \gamma ) dq$ is the conditional probability
that an oblate spheroid with an intrinsic short-to-long axis ratio
$\gamma$ has an observed apparent axis ratio in the range $q \to
q + dq$, averaged over all viewing angles.
The numerical value of $P_{\rm obl}$ is \citep{sa70}
\begin{equation}
P_{\rm obl} (q | \gamma ) = {q \over ( 1 - \gamma^2 )^{1/2} (q^2 - \gamma^2 )^{1/2} }
\end{equation}
if $\gamma \leq q \leq 1$, and $P_{\rm obl} = 0$ otherwise. Equation~(\ref{eq:volt})
is a Volterra equation of the first kind; in its discretized form, it can be
inverted by a process of forward substitution to find $N(\gamma)$ given
$f (q)$ (see \citet{vi05} for numerical details). 

In the bottom panel of Figure~\ref{fig:10}, the solid line
indicates the estimate of $N(\gamma)$ found by inverting the best
estimate of $f(q)$; the dashed lines are the 98\% confidence
intervals found from the inversion of the bootstrap estimates of
$f(q)$. The most probable value of $\gamma$, given the oblate hypothesis,
is $\gamma = 0.22$. For comparison to these $r$-band results, an
inclination-corrected sample of spiral galaxies from the 2MASS Large
Galaxy Atlas has a most probable thickness of $\gamma_B = 0.17$
in the $B$ band and $\gamma_K = 0.25$ in the $K_s$ band \citep{ry06}.
A noteworthy property of our estimate of $N(\gamma)$ is that it
is negative for large values of $\gamma$. The 98\% confidence
interval is negative for $\gamma \geq 0.89$; that is, fewer than
1\% of our bootstrap resamplings give $N (\gamma) > 0$ in this
interval. This unphysical results permits us to reject the hypothesis
of perfect oblateness at the 99\% (one-sided) confidence level.

We can approximate a galaxy in our sample not as an oblate spheroid,
but as a triaxial ellipsoid, with axis lengths $a \geq b \geq c$.
The shape can then be expressed in terms of two parameters, which
we choose to be the short-to-long axis ratio, $\gamma \equiv c / a$,
and the disk ellipticity, $\epsilon \equiv 1 - b/a$. Once we
permit non-zero values of $\epsilon$, we can no longer use the
observed distribution $f (q)$ to uniquely determine the intrinsic
distribution $N(\gamma,\epsilon)$. However, we still can do
parametric fits to the distribution of shapes. It is found that a
useful parameterization is $N(\gamma,\epsilon) = N_\gamma (\gamma)
N_\epsilon (\epsilon)$, with $N_\gamma$ being a Gaussian,
\begin{equation}
N_\gamma (\gamma) \propto \exp \left[ - {( \gamma - \mu_\gamma )^2
\over 2 \sigma_\gamma^2 } \right] ,
\label{eq:gamma}
\end{equation}
and $N_\epsilon$ being a log-normal distribution
\citep{an01,ry04,ry06},
\begin{equation}
N_\epsilon (\epsilon) \propto {1 \over \epsilon}
\exp \left[ - { (\ln \epsilon - \mu )^2 \over 2 \sigma^2 } \right] \ .
\label{eq:epsilon}
\end{equation}
The best values of $\mu_\gamma$, $\sigma_\gamma$, $\mu$, and $\sigma$
were determined by a $\chi^2$ fit to the binned distribution of $q$
for the $n = 16{,}155$ exponential galaxies in our inclination-corrected
sample. The bin width chosen was $dq = 0.01$. After selecting values of
$\mu_\gamma$, $\sigma_\gamma$, $\mu$, and $\gamma$, we randomly chose
$n$ values of $\gamma$ and of $\epsilon$ from the distributions of
equation~(\ref{eq:gamma}) and (\ref{eq:epsilon}). For each $(\gamma,\epsilon)$
pair, a random viewing angle was chosen, and the resulting apparent
axis ratio was computed \citep{bi78}. The model axis ratios were
then binned in the same way as the actual, observed axis ratios.
Repeating this procedure 400 times for each $(\mu_\gamma,\sigma_\gamma,
\mu,\sigma)$ set, we calculated the mean and standard deviation for
the number of model galaxies in each $q$ bin, and computed a $\chi^2$
probability for that particular set of parameters. The best fit we
found was $\mu_\gamma = 0.216$ and $\sigma_\gamma = 0.067$ for the
distribution of disk thicknesses, with $\mu = -2.56$ and $\sigma = 0.91$
for the distribution of the natural logarithm of the ellipticity. The
$\chi^2$ probability yielded by this set of parameters is $P \sim 2 \times
10^{-7}$. In the top panel of Figure~\ref{fig:10}, the dotted red line
shows the distribution of $q$ expected from this set of parameters,
smoothed with the same kernel width as the real data (solid black line).

The distribution of ellipticity of our best-fitting parametric model,
$\ln \epsilon = -2.56 \pm 0.91$, implies a modal ellipticity
$\epsilon_{\rm mod} = 0.033$, a median ellipticity $\epsilon_{\rm med} =
0.077$ and a mean ellipticity $\epsilon_{\rm ave} = 0.11$. This
distribution of thicknesses is consistent with the implied ellipticity
of an inclination-corrected sample of spiral galaxies from the
2MASS Large Galaxy Atlas \citep{ry06}, if the shape parameter for
the 2MASS spirals is the axis ratio of the 25 mag arcsec$^{-2}$
isophote. If the potential ellipticity equals the disk ellipticity,
then $\ln \epsilon = -2.56 \pm 0.91$ implies roughly one magnitude
of scatter in the Tully-Fisher relation \citep{tu77}.

\section{DISCUSSION}
\label{sec-disc}

We have selected a population of galaxies from the Sloan
Digital Sky Survey which are at low redshift ($z < 0.06$),
which are relatively luminous ($M_r \leq -19$), and which
are well described by an exponential surface brightness
profile (${\rm fracDeV} < 0.1$). As we have shown, the
properties of these galaxies are consistent with their
being a population of slightly elliptical disks containing
dust. The median dimensionless disk thickness for these
galaxies in the $r$ band is $\gamma \approx 0.22$; the median
disk ellipticity is $\epsilon \approx 0.08$.

By fitting the luminosity function for galaxies with different
apparent axis ratio $q$, we found that the apparent dimming
$\Delta M_r$ is not linearly proportional to $\log q$, but instead
is much better fitted by $\Delta M_r \propto ( \log q )^2$.
The dependence of dimming on inclination is a valuable clue
to the dust properties within disk galaxies. If certain
simplifying assumptions are made, the expected attenuation
as a function of inclination can be computed for model
galaxies. For instance, \citet{fe99} assumed that dust
had either the extinction curve found for Milky Way dust
or for Small Magellanic Cloud dust \citep{go97}. They
assumed that disks were perfectly axisymmetric, with
a horizontal scale length $r = 4 {\rm\,kpc}$ that was
the same for both stars and dust. The scale height
of the stars was assumed to be $z_\star = 0.35$, but
the scale height of the dust was allowed to vary.
\citet{fe99} found that dimming in the $B$ and $I$
bands were proportional to $\log q$ only when the
dust scale height was greater than that of the stars.
However, observation of nearby edge-on disk galaxies
\citep{xi99} indicates that the dust scale height
is about half the star scale height.

\citet{ro08}, using a Monte-Carlo radiative-transfer
code to make calculations of internal extinction in
dusty galaxies, found that a quadratic dependence of
dimming on $\log q$ provides a good fit for all
plausible dust scale heights, scale lengths, and
metallicity gradients. Since they were using hydrodynamic
galaxy models with spiral structure, they were
able to confirm that nonaxisymmetric structures such
as spiral arms did not significantly affect the
dependence of the total dimming $\Delta M$ on the
apparent axis ratio $q$. 

In our sample of exponential galaxies from the
Sloan Digital Sky Survey, once the absolute magnitude
of a galaxy is corrected for the inclination-dependent
dimming, the mean $u-r$ color observed is linearly
dependent on the corrected $M_r^f$. For nearly
face-on galaxies, with $q > 0.9$, the dependence
of $u-r$ on $M_r^f$ is relatively weak. We find
$b = -0.096$; that is, less than 0.1 magnitude of
reddening in $u-r$ for each magnitude brighter in
$M_r^f$. For the edge-on galaxies, with $q < 0.3$,
the dependence of $u-r$ on $M_r^f$ is much stronger,
with $b = -0.265$. The mean color -- absolute magnitude
dependence is a manifestation of the metallicity --
luminosity dependence. High-metallicity galaxies
have both redder stellar populations and higher
dust contents. For face-on exponential galaxies,
the dust effects are minimized, and we see the
effect of metallicity on stellar populations.
For edge-on exponential galaxies, we see, in
addition, the effect of metallicity on the
dust content. Exponential galaxies with
$M_r^f \sim -21.5$ have a typical color
$\langle u - r \rangle \sim 1.8$ when
seen face-on, placing them at the tip of
the blue sequence in a color --  magnitude
diagram. However, the same bright exponential
galaxies when seen edge-on will have
$M_r \sim -21$ and $\langle u -r \rangle
\sim 2.5$, a degree of reddening that smuggles
them into the red sequence, as usually defined.


Funding for the SDSS and SDSS-II has been provided
by the Alfred P. Sloan Foundation, the Participating Institutions,
the National Science Foundation,
the U.S. Department of Energy, the National Aeronautics
and Space Administration, the Japanese Monbukagakusho,
the Max Planck Society, and the Higher Education Funding Council
for England. The SDSS website is \url{http://www.sdss.org/}.
The SDSS is managed by the Astrophysical Research Consortium (ARC) for
the Participating Institutions. 
The Participating Institutions are the American Museum of Natural History,
Astrophysical Institute Potsdam, University of Basel,
University of Cambridge, Case Western Reserve University,
University of Chicago, Drexel University, Fermilab,
the Institute for Advanced Study, the Japan Participation Group,
Johns Hopkins University, the Joint Institute for Nuclear Astrophysics,
the Kavli Institute for Particle Astrophysics and Cosmology,
the Korean Scientist Group, the Chinese Academy of Sciences (LAMOST),
Los Alamos National Laboratory, the Max-Planck-Institute for
Astronomy (MPIA), the Max-Planck-Institute for Astrophysics (MPA),
New Mexico State University, The Ohio State University,
University of Pittsburgh, University of Portsmouth, Princeton University,
the United States Naval Observatory, and the University of Washington.

\include{tab1}


\end{document}

%% file: tab1.tex
\begin{deluxetable}{ccc}
\tablewidth{0pt}
\tablecaption{Mean color vs. face-on luminosity: $\mu (u-r) = a + b ( M_r^f + 20.5)$\label{tab:1}}
\tablehead{\colhead{q}& \colhead{a} & \colhead{b} }
\startdata
0.26 & 2.15 & -0.265 \\
0.35 & 2.05 & -0.210 \\
0.45 & 1.96 & -0.212 \\
0.55 & 1.90 & -0.184 \\
0.65 & 1.84 & -0.165 \\
0.75 & 1.81 & -0.162 \\
0.85 & 1.77 & -0.127 \\
0.95 & 1.75 & -0.096 \\
\enddata
\end{deluxetable}

%% file: ms.bbl
\newpage

\begin{thebibliography}{}

\bibitem[Abramson(1982)]{ab82}
Abramson, I. S. 1982, Ann. Stat., 10, 1217

\bibitem[Adelman-McCarthy et al.(2008)]{ad08}
Adelman-McCarthy, J. K.; for the SDSS Collaboration.
2008, AJ, submitted (astro-ph/0707.3413)

\bibitem[Alam \& Ryden(2002)]{al02}
Alam, S. M. K., \& Ryden, B. S. 2002, ApJ, 570, 610

\bibitem[Andersen et al.(2001)]{an01}
Andersen, D. R., Bershady, M. A., Sparke, L. S., Gallagher, J. S.,
\& Wilcots, E. M. 2001, ApJ, 551, L131

\bibitem[Baldry et al.(2004)]{ba04}
Baldry, I. K., Glazebrook, K., Brinkmann, J., Ivezic, Z.,
Lupton, R. H., Nichol, R. C., \& Szalay, A. S. 2004, ApJ, 600

\bibitem[Bell \& de Jong(2000)]{be00}
Bell, E. F., \& de Jong, R. S. 2000, MNRAS, 312, 497

\bibitem[Binney(1978)]{bi78}
Binney, J. 1978, MNRAS, 183, 779

\bibitem[Binney \& de Vaucouleurs(1981)]{bi81}
Binney, J., \& de Vaucouleurs, G.
1981, MNRAS, 194, 679

\bibitem[Blanton et al.(2003)]{bl03}
Blanton, M. R., et al. 2003, ApJ, 594, 186

\bibitem[Bottinelli et al.(1995)]{bo95}
Bottinelli, L., Gouguenheim, L., Paturel, G., \& Teerikorpi, P.
1995, A\&A, 296, 64

\bibitem[Chang et al.(2006)]{ch06}
Chang, R., Shen, S., Hou, J., Shu, C., \& Shao, Z.
2006, MNRAS, 372, 199

\bibitem[de Vaucouleurs(1948)]{dV48}
de Vaucouleurs, G. 1948, Ann. d'Astrophys. 11, 247

\bibitem[de Vaucouleurs et al.(1991)]{dV91}
de Vaucouleurs, G., de Vaucouleurs, A., Corwin, H. G., Jr.,
Buta, R. J., Paturel, G., \& Fouque, P.
1991, Third Reference Catalogue of Bright Galaxies
(New York: Springer-Verlag) (RC3)

\bibitem[Fasano et al.(1993)]{fa93}
Fasano, G., Amico, P., Bertola, F., Vio, R., \& Zeilinger, W. W.
1993, MNRAS, 262, 109

\bibitem[Ferrara et al.(1999)]{fe99}
Ferrara, A., Bianchi, S., Cimatti, A., \& Giovanardi, C.
1999, ApJS, 123, 437

\bibitem[Fukugita et al.(1995)]{fu95}
Fukugita, M., Shimasaku, K., \& Ichikawa, T.
1995, PASP, 107, 945


\bibitem[Gordon et al.(1997)]{go97}
Gordon, K. D., Calzetti, D., \& Witt, A. N.
1997, ApJ, 487, 625

\bibitem[Grosbol(1985)]{gr85}
Grosbol, P. J. 1985, A\&AS, 60, 261

\bibitem[Hubble(1926)]{hu26}
Hubble, E. P. 1926, ApJ, 64, 321

\bibitem[Lambas et al.(1992)]{la92}
Lambas, D. G., Maddox, S. J., \& Loveday, J.
1992, MNRAS, 258, 404

\bibitem[MacArthur et al.(2004)]{ma04}
MacArthur, L. A., Courteau, S., Bell, E., \& Holtzman, J. A.
2004, ApJS, 152, 175

\bibitem[Masters et al.(2003)]{ma03}
Masters, K. L., Giovanelli, R., \& Haynes, M. P.
2003, AJ, 126, 158

\bibitem[Rocha et al.(2008)]{ro08}
Rocha, M., Jonsson, P., Primack, J. R., \& Cox, T. J.
2008, MNRAS, in press (astro-ph/0702513v2)

\bibitem[Ryden(1992)]{ry92}
Ryden, B. S. 1992, ApJ, 396, 445

\bibitem[Ryden(2004)]{ry04}
Ryden, B. S. 2004, ApJ, 601, 214

\bibitem[Ryden(2006)]{ry06}
Ryden, B. S. 2006, ApJ, 641, 773

\bibitem[Sandage et al.(1970)]{sa70}
Sandage, A., Freeman, K. C., \& Stokes, N. R.
1970, ApJ, 160, 831

\bibitem[S\'ersic(1968)]{se68}
S\'ersic, J. L. 1968, Atlas de Galaxias Australes
(Cordoba: Obs. Astron.)

\bibitem[Shao et al.(2007)]{sh07}
Shao, Z., Xiao, Q., Shen, S., Mo, H. J., Xia, X., \& Deng, Z.
2007, ApJ, 659, 1159

\bibitem[Smith et al.(2002)]{sm02}
Smith, J. A., et al. 2002, AJ, 123, 2121


\bibitem[Strateva et al.(2001)]{st01}
Strateva, I. 2001, AJ, 122, 1861

\bibitem[Tremblay \& Merritt(1995)]{tr95}
Tremblay, B., \& Merritt, D. 1995, AJ, 110, 1039

\bibitem[Tully \& Fisher(1977)]{tu77}
Tully, R. B., \& Fisher, J. R. 1977, A\&A, 54, 661

\bibitem[Tully et al.(1998)]{tu98}
Tully, R. B., Pierce, M. J., Huang, J.-S., Saunders, W.,
Verheijen, M. A. W., \& Witchalls, P. L.
2998, AJ, 115, 2264

\bibitem[Vincent \& Ryden(2005)]{vi05}
Vincent, R. A., \& Ryden, B. S. 2005, ApJ, 623, 137

\bibitem[Vio et al.(1994)]{vi94}
Vio, R., Fasano, G., Lazzarin, M., \& Lessi, O. 1994, A\&A, 289, 640

\bibitem[Xilouris et al.(1999)]{xi99}
Xilouris, E. M., Byun, Y. I., Kylafis, N. D., Paleolougou, E. V.,
\& Papamastorakis, J. 1999, A\&A, 344, 868

\bibitem[York et al.(2000)]{yo00}
York, D. G., et al. 2000, AJ, 120, 1579

\end{thebibliography}
